\documentclass[journal]{IEEEtran}
\usepackage{cite}
\usepackage{amsmath,amssymb,amsfonts}
\usepackage{graphicx}

\usepackage{xurl}
\usepackage[ruled,vlined,linesnumbered]{algorithm2e}
\usepackage{booktabs}
\usepackage{siunitx}
\usepackage{xcolor}

\newtheorem{theorem}{Theorem}
\newtheorem{lemma}{Lemma}
\newtheorem{proposition}{Proposition}
\newtheorem{corollary}{Corollary}

\newtheorem{remark}{Remark}
\newtheorem{assumption}{Assumption}

\def\BibTeX{{\rm B\kern-.05em{\sc i\kern-.025em b}\kern-.08em
    T\kern-.1667em\lower.7ex\hbox{E}\kern-.125emX}}
\begin{document}
\title{Distributed Regenerative Resource Allocation with Exact Centralized Optimality}
\author{Rajul Kumar, Goutam Das, and Ningshi Yao, \IEEEmembership{Member, IEEE}
\thanks{This work was partially supported by NIH 1R01EB038710-02, ONR N00014-26-1-2153, and NSF CAREER 2539218.}
\thanks{Rajul Kumar and Ningshi Yao are with the Department of Electrical and Computer Engineering, George Mason University, Fairfax, VA, USA (e-mail: rkumar29@gmu.edu; nyao4@gmu.edu).}
\thanks{Goutam Das is with the School of Aeronautics and Astronautics, Purdue University, West Lafayette, IN, USA (e-mail: das347@purdue.edu).}

\thanks{\copyright~2026 IEEE. Personal use of this material is
permitted. Permission from IEEE must be obtained for all other
uses, in any current or future media, including
reprinting/republishing this material for advertising or
promotional purposes, creating new collective works, for resale
or redistribution to servers or lists, or reuse of any copyrighted
component of this work in other works.}
}
\maketitle
\thispagestyle{empty}


\begin{abstract}
This letter presents a Distributed Regenerative Resource Allocation (or DRRA) framework for multi-agent networks with utility-driven resource regeneration, in which the resource available for the next allocation round depends endogenously on the utility generated by the agents in the previous allocation. We develop fully distributed, two-stage
inner--outer dynamics that perform primal--dual allocation and resource
regeneration. Under sufficient distributed algorithmic gain conditions, the inner dynamics exhibit local exponential convergence to a Lyapunov-stable
equilibrium. After inner convergence, the DRRA allocation exactly recovers the centralized optimum, and the local resource-reference updates
exactly reproduce the centralized resource-regeneration dynamics.
\end{abstract}

\begin{IEEEkeywords}
Distributed control, Networked control systems, Optimization algorithms, Cooperative control
\end{IEEEkeywords}


\section{Introduction} \label{sec1}
Distributed resource allocation (or DRA) is fundamental in multi-agent networks, where resources are spatially dispersed, and each agent controls its local resources, generating utility upon task completion. By avoiding a central coordinator, distributed approaches eliminate single points of failure, thereby improving reliability. Practical DRA applications include energy management in
microgrids, workload allocation across data centers, and
allocation of sensing and computing resources in multi-robot
networks\cite{doostmohammadian2025survey}.
In regenerative settings, the utility generated by current
allocations can replenish the resource pool for subsequent
rounds. For example, profits from microgrid energy trading
or computing services can be reinvested to increase the
operating budget available for the next allocation round. Consequently, the utility generated from current allocations determines the additional resource available in the next round, closing the feedback loop between agents' task performance and resource allocation. However, limited work models what occurs after agents utilize these resources and generate utility, or how this utility contributes to successive allocation rounds and evolves in a distributed manner over time while achieving centralized optimality.

Many studies have rigorously investigated time-varying resource allocation in different contexts. For example, although the resource is termed time-varying in\cite{deng2026distributed, yue2025predefined, ding2024distributed, wang2020finite, xu2025feedback}, its evolution is externally prescribed rather than regenerated from the utility earned by agents. Consequently, there is no utility-driven feedback, future resource availability does not depend on current allocation, and successive allocation rounds remain uncoupled. In\cite{wang2020continuous}, the cost functions and resource availability vary according to externally specified time-dependent signals. Similarly,\cite{wang2021asynchronous} addresses DRA with asynchronous updates and a changing network topology but lacks resource regeneration across allocation rounds. For externally specified resource, the authors in\cite{shi2023finite, luan2025fully} provide finite- and prescribed-time convergence, respectively. Similarly, the work in\cite{wang2022quadratic} considers distributed resource allocation with exact tracking of the centralized optimum and establishes convergence using Karush-Kuhn-Tucker (KKT) conditions under time-varying optimization. However, the variation is driven by exogenous signals independent of the optimization process; thus, the distributed algorithm only tracks an exogenous moving optimum without intertemporal resource coupling. 

In\cite{doan2020uncertainty}, the authors consider DRA under uncertain resource information and provide convergence guarantees in expectation. The authors in\cite{yao2025human} address human-in-the-loop distributed optimization but do not consider regenerative resource dynamics. The studies in\cite{zhou2026nash, zhao2026unknown, huo2025private} consider DRA without resource regeneration. The work in\cite{zhou2026nash} seeks a Nash equilibrium rather than a centralized optimum, while\cite{zhao2026unknown, huo2025private} provide approximate convergence guarantees. The closest related study,\cite{tao2025recycling}, introduces centralized budget recycling, in which unused budget is returned for future allocations. However, it relies on a centralized framework and provides neither a distributed
regeneration mechanism nor theoretical guarantees for exact distributed recovery of the centralized optimum. Thus,
a utility-driven distributed resource regeneration mechanism
for multi-agent networks with exact centralized optimality
remains undiscovered, and a rigorous analysis of such a
mechanism remains an open problem.

\textit{
Addressing this research gap, as the main contribution of this paper,}
we propose a novel Distributed Regenerative Resource Allocation (or DRRA) framework, in which the total resource available for the next allocation round is endogenously regenerated from the utility generated by the agents. Consequently, through a utility-driven closed-loop feedback mechanism, future resource availability depends directly on previous allocations. We develop a fully distributed resource allocation and regeneration algorithm with rigorous theoretical guarantees, in which agents use only neighbor-to-neighbor communication to recover the centralized optimal allocation. We present two-stage inner--outer distributed dynamics, in which the inner dynamics attain an optimal, resource-feasible equilibrium, while the outer dynamics regenerate the resource for the next allocation round. Under sufficient distributed algorithmic gain conditions, the inner dynamics exhibit local exponential convergence to a
Lyapunov-stable equilibrium. After inner convergence, local reference updates exactly recover centralized resource regeneration, with simulations supporting the theoretical results.
 

\textbf{Notation:} $\mathbb{R}_{\geq 0}$, $\mathbb{R}_{>0}$, and $\mathbb{Z}_{\ge 0}$ denote the sets of nonnegative reals, strictly positive reals, and nonnegative integers, respectively. $\mathbf{1}\!\in\!\mathbb{R}^n$ and $L\!\in\!\mathbb{R}^{n\times n}$ denote the all-ones vector and graph Laplacian, respectively. Equilibria and optimal values are denoted by an asterisk and superscript ``$\mathrm{o}$'', respectively.


\section{Problem Formulation and CRRA Benchmark}\label{sec2}
Consider a multi-agent system consisting of $n\!\geq\!2$ agents, indexed by the set $\mathcal{V}\! =\! \{1,\dots,n\}$. At each discrete time step $T \!\in\! \mathbb{Z}_{\geq 0}$, agent $i \!\in\! \mathcal{V}$ is assigned a local resource $b_i(T)\! \in \!\mathbb{R}_{\geq 0}$ to execute an independent task. Let $\mathbf{b}(T)\! =\! [b_1(T),\! \dots\!, b_n(T)]^\top\! \in\! \mathbb{R}_{\geq 0}^n$ denote the global resource-allocation vector. Utilizing its assigned resource, agent $i$ generates utility $U_i(b_i(T))$. The nonlinear utility function $U_i:\mathbb{R}_{\geq0}\!\to\!\mathbb{R}$ is known locally to agent $i$ and depends solely on its allocation $b_i(T)$. It is considered over the interior operating region specified in Assumption~\ref{ass1}. The corresponding marginal utility is $\frac{dU_i(b_i(T))}{db_i(T)}$. For a given total resource $B(T)\! \in\! \mathbb{R}_{>0}$ at time $T$, the allocation problem is formulated as
\begin{equation}
\label{eq1}
\mathbf{b}^{\mathrm{o}}(T)
\!=\!
\operatorname*{arg\,max}_{\mathbf{b}(T) \in \mathbb{R}_{\geq 0}^{n}}
\;
\sum_{i=1}^{n} U_i(b_i(T))
\;
\text{s.t.}
\;
\sum_{i=1}^{n} b_i(T)\! = \!B(T).
\end{equation}
For each round $T$, problem~\eqref{eq1} determines the optimal allocation $\mathbf{b}^{\mathrm{o}}(T)$ that maximizes the aggregate utility while ensuring that the total assigned resource exactly equals the available capacity $B(T)$. Evaluating \eqref{eq1} at the optimum $\mathbf{b}^{\mathrm{o}}(T)$ yields the maximized aggregate utility $\sum_{i=1}^{n} U_i\!\left(b_i^{\mathrm{o}}(T)\right)$. A fraction of this generated utility, scaled by a regeneration factor $\epsilon \in (0,1)$, is reinjected into the total resource pool. The factor $\epsilon$ determines the additional
resource obtained per unit of generated utility. For example, when resource and utility represent monetary
budget and profit, respectively, $\epsilon$ reflects a practical
budgeting decision, i.e., the fraction of profit reinvested in the
next round. Across successive rounds, the total resource evolves according to
\begin{equation}
\label{eq2}
B(T+1)
=
B(T)
+
\epsilon
\sum_{i=1}^{n}
U_i\!\left(b_i^{\mathrm{o}}(T)\right).
\end{equation}
While \eqref{eq1} is static within each round, its solution
determines the next round's resource and feasible set through
\eqref{eq2}. Reproducing both the exact centralized optimal allocation
and its induced resource dynamics in a fully distributed
manner using only local utilities and neighbor exchanges
is challenging due to the dependence of future resource
constraints on current allocations: errors in one round
can change subsequent feasible sets, requiring consistent
guarantees for both allocation and regeneration.

\begin{assumption}\label{ass1}
For each round $T$ considered, the centralized optimizer
$\mathbf b^{\mathrm o}(T)$ lies in the interior of a compact
operating set
$\mathcal B_{\mathrm{op}}\subset\mathbb R_{\geq0}^{n}$,
so $b_i^{\mathrm o}(T)>0$ for every agent $i$.
Each utility $U_i$ is twice continuously differentiable
and satisfies $U_i''(b_i)<0$ for all $b_i\in[0,B(T)]$,
ensuring strict concavity.
\end{assumption}

Subsection~\ref{sec2.1} establishes the benchmark Centralized Regenerative Resource Allocation (or CRRA) framework.

\subsection{Benchmark: Centralized Regenerative Resource Allocation (CRRA)}\label{sec2.1}
For a fixed allocation round $T$, the centralized optimal solution is derived via the Lagrangian of \eqref{eq1}, defined as
\begin{equation}
\label{eq3}
\mathcal{L}_T(\mathbf{b},\delta)
=
\sum_{i=1}^{n} U_i(b_i(T))
-
\delta
\Bigl(
\sum_{i=1}^{n} b_i(T)
-
B(T)
\Bigr),
\end{equation}
where $\delta \! \in \! \mathbb{R}$ is the associated Lagrange multiplier. 

The following proposition establishes the equal-marginal-utility condition that characterizes this optimal CRRA solution.


\begin{proposition}
\label{pro1}
Under the CRRA framework and Assumption~\ref{ass1}, any optimal solution $\mathbf{b}^{\mathrm{o}}(T)$ of \eqref{eq1} satisfies the KKT stationarity condition \eqref{eq4}, which requires all agents to have identical marginal utilities:
\begin{equation}
\label{eq4}
\left. \frac{dU_i(b_i(T))}{db_i(T)} \right|_{b_i(T)=b_i^{\mathrm{o}}(T)}
=
\delta^{\mathrm{o}}(T),
\qquad
\forall i \in \mathcal{V},
\end{equation}
where $\delta^{\mathrm{o}}(T)$ is the optimal dual variable associated with the global resource constraint.
\end{proposition}

\begin{IEEEproof}
Fix a regeneration round $T$ and the corresponding total resource $B(T)$.
Differentiating \eqref{eq3} with respect to $b_i(T)$ gives
$\frac{\partial \mathcal L_T}{\partial b_i(T)}
=
\frac{dU_i(b_i(T))}{db_i(T)}-\delta$.
By Assumption~\ref{ass1}, $\mathbf b^{\mathrm{o}}(T)$ is interior; hence the
nonnegativity constraints are inactive. Therefore, the KKT stationarity
condition gives
$\left.
{\partial \mathcal L_T}/{\partial b_i(T)}
\right|_{b_i(T)=b_i^{\mathrm{o}}(T)}
=0$, and thus, with $\delta=\delta^{\mathrm{o}}(T)$ at the optimum, \eqref{eq4} follows. Moreover, since $\mathbf b^{\mathrm{o}}(T)$ solves \eqref{eq1}, it satisfies
$\sum_{i=1}^{n}b_i^{\mathrm{o}}(T)=B(T)$. Hence,
$\mathbf b^{\mathrm{o}}(T)$ satisfies the KKT stationarity condition and the
resource constraint. Strict concavity of the objective and convexity of the
constraint set imply that this optimizer is unique.
\end{IEEEproof}


Condition \eqref{eq4} in Proposition~\ref{pro1} implies that at the optimal allocation, all agents have equal marginal utilities. Hence, any infinitesimal feasible reallocation of the resource produces no first-order improvement in the aggregate utility. 

Having established the CRRA benchmark and its KKT conditions, we now consider a distributed framework without a central coordinator. Agents communicate only with their immediate neighbors over a graph $\mathcal G\!=\!(\mathcal V,\mathcal E)$, characterized in Assumption~\ref{ass2}, with $\mathcal N_i$ denoting the neighbor set of agent $i$.

\begin{assumption}
\label{ass2}
The communication graph $\mathcal{G}$ is fixed, strongly connected, and weight-balanced. Accordingly, its Laplacian $L$ satisfies $L \mathbf{1} = \mathbf{0}$ and $\mathbf{1}^\top L = \mathbf{0}^\top$.
\end{assumption}

\textit{\textbf{Problem Statement:}}
The objective is to design a fully distributed allocation framework in which agents exchange information exclusively with immediate neighbors over a graph satisfying Assumption~\ref{ass2}. At each round $T$, the distributed protocol must recover the centralized optimal allocation of Subsection~\ref{sec2.1} by satisfying the KKT conditions \eqref{eq4} and the global resource constraint \eqref{eq1}, while preserving the exact regeneration dynamics \eqref{eq2}.

Addressing the above problem statement, Section~\ref{sec3} presents the proposed Distributed Regenerative Resource Allocation (or DRRA) framework and its main theoretical results.


\section{Main Results: Distributed Regenerative Resource Allocation (DRRA)}\label{sec3}

The DRRA framework is formulated over a dual-time-scale
architecture as a two-stage inner--outer protocol. The discrete index $T \!\in\! \mathbb{Z}_{\geq 0}$ denotes the outer resource-regeneration round, while $t \!\in \!\mathbb{Z}_{\geq 0}$ denotes the inner distributed iteration. \\
\textbf{Step 0 (Initialization):} At each regeneration round $T$, every agent $i \in \mathcal{V}$ maintains a local resource reference $c_i(T) \in \mathbb{R}_{>0}$ satisfying $\sum_{i=1}^{n} c_i(T) = B(T)$. A natural initialization is $c_i(0) = B(0)/n$ for all $i \in \mathcal{V}$. For a fixed outer round $T$, this reference $c_i(T)$ is held constant while agent $i$ iteratively updates its optimal allocation estimate $b_i(t;T)$, an auxiliary coordination state $z_i(t;T)$, and a local dual variable $\lambda_i(t;T)$ on the inner time scale $t$.\\
\textbf{Step 1:} At each inner iteration $t$, agent $i$ receives
three scalar quantities from each immediate neighbor
$j\in\mathcal N_i$: the marginal utility
$\frac{dU_j(b_j(t;T))}{db_j}$, auxiliary state $z_j(t;T)$, and
dual variable $\lambda_j(t;T)$. Using these quantities and its
local states, agent $i$ updates the inner dynamics
\eqref{eq5}--\eqref{eq7} according to
\begin{align}
b_i(t\!+\!1;T)
&=\!
b_i(t;T)
\!\!+\!\!
\gamma
\frac{dU_i(b_i(t;\!T))}{db_i}
\!\!+\!\!
k\!\!\sum_{j\in\mathcal N_i}\!\!a_{ij}
\frac{dU_j(b_j(t;\!T))}{db_j}
\notag
\\[-1mm]
&\quad
\!-\!
k\!\!\sum_{j\in\mathcal N_i}\!\!a_{ij}
\frac{dU_i(b_i(t;T))}{db_i}
\!-\!
\alpha\lambda_i(t;T),
\label{eq5}
\end{align}
\begin{align}
z_i(t\!+\!1;T)
&=\!
z_i(t;T)
\!-\!
\eta
\!\!\sum_{j\in\mathcal N_i}\!\!a_{ij}
\!\left(
\lambda_i(t;T)
\!-\!
\lambda_j(t;T)
\right),
\label{eq6}
\end{align}
\begin{align}
\lambda_i(t\!+\!1;T)
&=\!
\lambda_i(t;T)
+
\beta \; b_i(t;T)
+
\beta \sum_{j\in\mathcal N_i} a_{ij}z_i(t;T)
\notag
\\[-1mm]
&\quad
-
\beta \sum_{j\in\mathcal N_i} a_{ij}z_j(t;T)
-
\beta \; c_i(T),
\label{eq7}
\end{align}
where $\gamma,k,\alpha,\eta,\beta\!>\!0$ are distributed gains and
$a_{ij}\geq 0$ is the communication weight associated with the edge
$(j,i)\in\mathcal E$. The auxiliary state $z_i$ accumulates neighboring dual-variable
differences. Through the dual dynamics \eqref{eq7}, it allows
agents' allocations to differ from their local resource
references while helping enforce the total resource constraint
at equilibrium.

In \eqref{eq5}, the gradient and neighbor-coupling terms seek the equal-marginal-utility condition in Proposition~\ref{pro1}. And the dual feedback term $-\alpha\lambda_i(t;T)$, together with the auxiliary \eqref{eq6} and dual dynamics \eqref{eq7}, enforces the global resource constraint \eqref{eq1}, thereby yielding a distributed primal--dual mechanism.

Let $(\mathbf b^\star(T),\mathbf z^\star(T),
\boldsymbol\lambda^\star(T))$ be an equilibrium of
\eqref{eq5}--\eqref{eq7}. Section~\ref{subsec_sp} establishes
local convergence to this equilibrium under distributed gain
conditions, and Section \ref{sec_3_b} (Theorem~\ref{thm1}) establishes exact recovery of the centralized allocation.

\textbf{Step 2:} After convergence of the inner dynamics \eqref{eq5}--\eqref{eq7}, agent $i$ evaluates $U_i(b_i^\star(T))$ and updates its local resource reference at the outer regeneration round $T\!+\!1$ as
\begin{equation}
\label{eq8}
c_i(T+1)
=
c_i(T)
+
\epsilon U_i(b_i^\star(T)),
\qquad
\forall i\in\mathcal V.
\end{equation}
Note that only the sum of the regenerated utilities is required to reproduce \eqref{eq2}; thus, the regenerated utilities, and hence the individual local reference increments, may differ across agents.
To facilitate reproducibility, we provide pseudo-code summarizing the DRRA in Algorithm \ref{alg1}.

\begin{algorithm}[t]
\fontsize{10pt}{12pt}\selectfont
\caption{Pseudo-code DRRA (Step 0-2)}
\label{alg1}

\textbf{Given:} Network $\mathcal G$ and weights $a_{ij}$
satisfying Assumption~\ref{ass2}, utilities $U_i$
satisfying Assumption~\ref{ass1}, and gains
$\gamma,k,\alpha,\eta,\beta$ satisfying
Proposition~\ref{pro2}\;

\textbf{Initialization:} $T\gets0$,
$c_i(0)\gets B(0)/n$, $\forall i\in\mathcal V$\;

    \While{outer termination criterion is not met}{
    Hold $c_i(T)$ fixed, $\forall i\in\mathcal V$\;
    Initialize $b_i(0;T)$, $z_i(0;T)$, and
    $\lambda_i(0;T)$ under the conditions of
    Proposition~\ref{pro2}; set $t\gets0$\;

    \Repeat{inner convergence}{
        Each agent $i$ computes $U_i'(b_i(t;T))$\;
        Each agent $i$ sends $U_i'(b_i(t;T))$,
        $z_i(t;T)$, and $\lambda_i(t;T)$ to its
        outgoing neighbors\;
        Each agent $i$ receives $U_j'(b_j(t;T))$, $z_j(t;T)$,
        and $\lambda_j(t;T)$ from each neighbor $j\in\mathcal N_i$\;
        Each agent $i$ updates $b_i(t+1;T)$,
        $z_i(t+1;T)$, and $\lambda_i(t+1;T)$
        via (5)--(7), using
        iteration-$t$ values\;
        $t\gets t+1$\;
    }

    Obtain the limiting allocation
    $b_i^\star(T):=\lim_{t\to\infty}b_i(t;T)$,
    $\forall i\in\mathcal V$\;
    Each agent $i$ updates
    $c_i(T+1)\gets c_i(T)+\epsilon U_i(b_i^\star(T))$
    via (8)\;
    $T\gets T+1$\;
}

\textbf{Return:} Converged allocations and updated
references\;
\end{algorithm}

\begin{remark}[Computational Complexity of DRRA]
Each inner DRRA iteration \eqref{eq5}--\eqref{eq7} requires
$O(1+|\mathcal N_i|)$ computation per agent, growing with its
number of incoming neighbors $|\mathcal N_i|$. Across the
network, the total computation is $O(n+|\mathcal E|)$, where
$n$ is the number of agents and $|\mathcal E|$ is the number
of directed communication links. Thus, the per-iteration
complexity is polynomial, with constant per-agent and linear
total cost for uniformly bounded degrees. The outer update
\eqref{eq8} requires $O(1)$ computation per agent.
\end{remark}

Compared with\cite{tao2025recycling}, DRRA requires neither Gaussian-process
model fitting nor acquisition optimization. For scalar allocations,
its per-agent computational complexity of
$O(1+|\mathcal N_i|)$ per inner iteration matches that of\cite{doan2020uncertainty, huo2025private} when their local minimizations have constant
computational cost.

Additionally, DRRA provides distributed utility-driven regeneration
and guarantees resource-demand balance at equilibrium over fixed,
strongly connected, weight-balanced networks. By comparison,\cite{doostmohammadian2025resilient} addresses uniformly connected
networks, heterogeneous time-varying delays, and possible nonlinearity
due to quantization, while ensuring resource-demand balance at every iteration
from feasible initialization. In a related study also considering strongly connected,
weight-balanced digraphs\cite{deng2018distributed}, agents
allocate fixed resources by exchanging dual and auxiliary
variables without sharing gradients. DRRA uses neighboring
marginal utilities in discrete-time primal--dual updates and
adds utility-driven regeneration after inner convergence,
linking current allocations to the resources available in
the next round.

\subsection{Convergence of the Inner DRRA Dynamics}\label{subsec_sp}
In this subsection, we establish the existence of an equilibrium of the inner dynamics \eqref{eq5}--\eqref{eq7} and derive conditions that guarantee local exponential convergence to it while the allocations remain in the positive interior operating region.

Let
$\mathbf g(\mathbf b)\!:=\![U_1'(b_1),\ldots,U_n'(b_n)]^\top$,
with $U_i'(b_i)\!:=\!\frac{dU_i(b_i)}{db_i}$, and
$\mathbf c(T)\!:=\![c_1(T),\ldots,c_n(T)]^\top$.
Stacking \eqref{eq5}--\eqref{eq7} gives
\begin{subequations}\label{eq9}
\begin{align}
\mathbf b(t+1;T)\!
&=\!\mathbf b(t;T)\!+\!(\gamma I\!-\!kL)\mathbf g(\mathbf b(t;T))
\!-\!\alpha\boldsymbol\lambda(t;T),\!\!
\label{eq9a}\\
\mathbf z(t+1;T)\!
&=\!\mathbf z(t;T)-\eta L\boldsymbol\lambda(t;T),
\label{eq9b}\\
\boldsymbol\lambda(t+1;T)
&=\boldsymbol\lambda(t;T)
\!+\!\beta(\mathbf b(t;T)\!+\!L\mathbf z(t;T)\!-\!\mathbf c(T)).\!
\label{eq9c}
\end{align}
\end{subequations}
\begin{remark}\label{rem1}
Assumption~\ref{ass2} and \eqref{eq6} imply
$\mathbf 1^\top\mathbf z(t+1;T)\!=\!\mathbf 1^\top\mathbf z(t;T)$;
thus, the average of $\mathbf z$ is invariant. Moreover,
$L(\mathbf z+a\mathbf 1)\!=\!L\mathbf z$ for any $a\in\mathbb R$, so a
common offset in $\mathbf z$ leaves \eqref{eq9c} unchanged and yields
an invariant direction in the Jacobian of \eqref{eq9}. We therefore
analyze the disagreement component $\Pi\mathbf z$, where
$\Pi\!:=\!I-\frac{1}{n}\mathbf 1\mathbf 1^\top$. Let
$Q\!\in\!\mathbb R^{n\times(n-1)}$ satisfy
$Q^\top Q\!=\!I_{n-1}$ and $QQ^\top\!=\!\Pi$, and define the reduced inner
state as
$\widetilde{\mathbf x}(t;T)\!:=\![
\mathbf b^\top(t;T),
(Q^\top\mathbf z(t;T))^\top,
\boldsymbol\lambda^\top(t;T)]^\top$.
Because the average of $\mathbf z$ is constant, convergence of
$\widetilde{\mathbf x}(t;T)$ implies convergence of the full state
$(\mathbf b,\mathbf z,\boldsymbol\lambda)$ in \eqref{eq9}.
The matrix $Q$ is used only for convergence analysis and need not be computed
by agents implementing DRRA in \eqref{eq5}--\eqref{eq7}.
\end{remark}

For a fixed round $T$, Proposition~\ref{pro2}
establishes equilibrium existence and conditions for local
exponential convergence and Lyapunov stability.
For an equilibrium allocation $\mathbf b^\star(T)$, define
$D(T):=\operatorname{diag}(
-U_1''(b_1^\star(T)),\ldots,-U_n''(b_n^\star(T)))$. The corresponding reduced Jacobian of \eqref{eq9} (Remark~\ref{rem1}) is
{\small
\begin{equation}\label{eq10}
J(T)=
\begin{bmatrix}
I-(\gamma I-kL)D(T) & 0 & -\alpha I\\
0 & I_{n-1} & -\eta Q^\top L\\
\beta I & \beta LQ & I
\end{bmatrix}.
\end{equation}
}
\begin{proposition}\label{pro2}
Under Assumptions~\ref{ass1} and~\ref{ass2}, fix a round
$T$ and hold $\mathbf c(T)$ constant, with
$\mathbf1^\top\mathbf c(T)\!=\!B(T)$.
Then the inner dynamics \eqref{eq5}--\eqref{eq7},
equivalently \eqref{eq9}, admit an equilibrium
$(\mathbf b^\star(T),\mathbf z^\star(T),
\boldsymbol\lambda^\star(T))$ with
$\mathbf b^\star(T)$ in the positive interior operating region
and $\mathbf1^\top\mathbf z^\star(T)
\!=\!\mathbf1^\top\mathbf z(0;T)$.

For distributed gains satisfying
$\max_{\nu\in\sigma(J(T))}|\nu|<1$,
the reduced Jacobian $J(T)$ in \eqref{eq10}, evaluated at
this equilibrium, is Schur stable, where $\sigma(\cdot)$
denotes the eigenvalues spectrum. Under this condition, there exists a symmetric
matrix $P_T\succ0$ satisfying
$J(T)^\top P_TJ(T)-P_T\prec0$, which is a linear matrix
inequality (LMI) in $P_T$ for prescribed gains.
There exist $q_T\in(0,1)$ and $r_T>0$ such that, if
$\|\widetilde{\mathbf x}(0;T)
-\widetilde{\mathbf x}^\star(T)\|_{P_T}\leq r_T$,
then the inner allocations remain in the positive interior
operating region and
\begin{equation*}
\|\widetilde{\mathbf x}(t;T)
-\widetilde{\mathbf x}^\star(T)\|_{P_T}
\leq q_T^t
\|\widetilde{\mathbf x}(0;T)
-\widetilde{\mathbf x}^\star(T)\|_{P_T},
\; t\geq0,
\end{equation*}
where $\|v\|_{P_T}:=(v^\top P_Tv)^{1/2}$ denotes the norm
induced by $P_T$.
Consequently, the reduced equilibrium is locally exponentially
stable and hence Lyapunov stable. The full equilibrium has
these stability properties relative to the invariant set
$\{(\mathbf b,\mathbf z,\boldsymbol\lambda):
\mathbf1^\top\mathbf z=\mathbf1^\top\mathbf z^\star(T)\}$ (Remark~\ref{rem1}).
\end{proposition}

\begin{IEEEproof}
We first derive sufficient conditions on the gains
$\gamma,k,\alpha,\eta,\beta$ ensuring Schur stability of $J(T)$.

For a fixed round $T$, write $D:=D(T)$ and $D\succ0$ (Assumption 1).
Decompose the network Laplacian as $L=S+K$, where
$S:=(L+L^\top)/2$ and $K:=(L-L^\top)/2$. Since the
graph is strongly connected and weight-balanced (Assumption 2), $S$ is a connected
undirected Laplacian, and hence $s_2:=\lambda_2(S)>0$. Also, let
$d_-:=\min_iD_{ii}>0$ and $d_+:=\max_iD_{ii}$.

For notational convenience, define
$\omega:=1/(2\|S\|_2)$,
$\vartheta:=\min\{1,d_-/(1+\frac{2}{3}d_+^2)\}$,
$p:=1+\frac{\sqrt{5}}{4}\vartheta$, and
$\varpi:=\min\{d_-,3\vartheta/8,\vartheta\omega^2s_2^2\}$.
The admissible graph-asymmetry margin is
$\zeta:=\varpi/(4p)-\omega\|K\|_2$, and we require $\zeta>0$.
For an undirected graph, $K=0$, so this condition holds
automatically. Finally, define
$\tau:=\min\{\frac12,
\frac{\zeta}{
2(1+\sqrt{1+\omega^2\|L\|_2^2}+\omega\|L\|_2)}\}$. The explicit sufficient distributed gain conditions are
\begin{subequations}\label{eqa1}
\begin{gather}
0<\gamma<
\frac{\varpi}{
2p\left(d_++\frac{\sqrt{5}}{2}+\frac{\varpi}{4p}\right)^2},
\label{eqa1a}\\
0<k<
\gamma\min\left\{
\frac{1}{\|L\|_2},
\frac{\zeta}{2\|LD\|_2}
\right\},
\label{eqa1b}\\
(1-\tau)\gamma<\alpha<(1+\tau)\gamma,
\label{eqa1c}\\
(1-\tau)\gamma<\beta<(1+\tau)\gamma,
\label{eqa1d}\\
(1-\tau)\omega^2\gamma<\eta<(1+\tau)\omega^2\gamma.
\label{eqa1e}
\end{gather}
\end{subequations}
We now prove that these conditions ensure Schur stability of
$J(T)$ in (10). First, rescale the reduced auxiliary coordinate using
$\mathcal T:=\operatorname{diag}
(I_n,\omega^{-1}I_{n-1},I_n)$. For notational convenience, define the gain
ratios $\kappa:=k/\gamma$, $a_\alpha:=\alpha/\gamma$,
$a_\beta:=\beta/\gamma$, and
$a_\eta:=\eta/(\omega^2\gamma)$. Direct substitution into
(10) gives
\begin{subequations}\label{eqA1}
\begin{equation}\label{eqA1a}
\widehat J:=\mathcal TJ(T)\mathcal T^{-1}=I+\gamma A,
\end{equation}
{\small
\begin{equation}\label{eqA1b}
A=
\begin{bmatrix}
-D+\kappa LD & 0 & -a_\alpha I\\
0 & 0 & -\omega a_\eta Q^\top L\\
a_\beta I & \omega a_\beta LQ & 0
\end{bmatrix}.
\end{equation}
}
\end{subequations}
Since $\widehat J=\mathcal TJ(T)\mathcal T^{-1}$,
$\widehat J$ is similar to $J(T)$, so both matrices have the same
eigenvalues. Therefore, it is sufficient to prove that
$\widehat J$ is Schur stable. Next, let $C:=\omega SQ$ and define the nominal  matrix associated with $S$
$A_s:=
[
[-D\;\;0\;\;-I]^\top\;\;
[0\;\;0\;\;-C^\top]^\top\;\;
[I\;\;C\;\;0]^\top
]^\top$.

From $\omega=1/(2\|S\|_2)$, we obtain
$\|C\|_2=\omega\|SQ\|_2\leq\omega\|S\|_2=1/2$.
Also, for any $w\in\mathbb R^{n-1}$, $Qw\in\mathbf1^\perp$ and
$\|Qw\|_2=\|w\|_2$ because $Q^\top Q=I_{n-1}$. Since $s_2$ is the
smallest positive eigenvalue of $S$, it follows that
$\|SQw\|_2\geq s_2\|Qw\|_2=s_2\|w\|_2$, and hence
$\|Cw\|_2\geq\omega s_2\|w\|_2$.

To establish strict decay for $A_s$, define $P:=I+\frac{\vartheta}{2}
[
[0\;\;0\;\;I]^\top\;\;
[0\;\;0\;\;-C^\top]^\top\;\;
[I\;\;-C\;\;0]^\top
]^\top$. The block matrix multiplying ${\vartheta}/{2}$ in the definition of $P$ has norm
$\sqrt{\lambda_{\max}(I+CC^\top)}
\leq\sqrt{1+\|C\|_2^2}\leq\sqrt{5}/2$.
Therefore, $(1-\frac{\sqrt{5}}{4}\vartheta)I
\preceq P\preceq pI$. Since $0<\vartheta\leq1$, it follows that $P\succ0$. We now evaluate the decay generated by $A_s$. For
$y=[u^\top,w^\top,v^\top]^\top$, direct multiplication gives
\begin{align}
&y^\top(A_s^\top P+PA_s)y
=-2u^\top Du+\vartheta\|u\|_2^2
-\vartheta u^\top Dv
\notag\\
&\qquad-\vartheta\|v\|_2^2
+\vartheta\|C^\top v\|_2^2
-\vartheta\|Cw\|_2^2.
\label{eqA3}
\end{align}
Since $D\succ0$ is diagonal,
$u^\top Du\geq d_-\|u\|_2^2$ and
$\|Dv\|_2\leq d_+\|v\|_2$. Young's inequality gives
$|u^\top Dv|\leq d_+\|u\|_2\|v\|_2
\leq\frac{2d_+^2}{3}\|u\|_2^2+\frac{3}{8}\|v\|_2^2$.
Furthermore, $\|C\|_2\leq1/2$ gives
$\|C^\top v\|_2^2\leq\frac14\|v\|_2^2$, while
$\|Cw\|_2\geq\omega s_2\|w\|_2$ gives
$\|Cw\|_2^2\geq\omega^2s_2^2\|w\|_2^2$.
Substituting these bounds into
\eqref{eqA3} yields
\begin{align}
y^\top(A_s^\top P+PA_s)y
\leq{}&
-\left[
2d_--\vartheta\left(1+\frac{2d_+^2}{3}\right)
\right]\|u\|_2^2
\notag\\
&-\frac{3\vartheta}{8}\|v\|_2^2
-\vartheta\omega^2s_2^2\|w\|_2^2.
\label{eq:a4}
\end{align}
From the definition of $\vartheta$, we have
$\vartheta(1+\frac{2d_+^2}{3})\leq d_-$, and therefore
$2d_--\vartheta(1+\frac{2d_+^2}{3})\geq d_-$.
Since
$\varpi=\min\{d_-,3\vartheta/8,\vartheta\omega^2s_2^2\}$,
each coefficient in \eqref{eq:a4} is at least
$\varpi$. Hence,
$y^\top(A_s^\top P+PA_s)y\leq-\varpi\|y\|_2^2$ for every $y$,
which is equivalent to $A_s^\top P+PA_s\preceq-\varpi I$. Next, write $A=A_s+E$. Using $L=S+K$ and directly comparing
the corresponding blocks of $A$ and $A_s$ gives
\begin{align}
&\|E\|_2
\leq \omega\|K\|_2+\kappa\|LD\|_2+|a_\alpha-1|
\notag\\
&\qquad+\sqrt{1+\omega^2\|L\|_2^2}|a_\beta-1|
+\omega\|L\|_2|a_\eta-1|.
\label{eq:E-bound}
\end{align}

The gain conditions \eqref{eqa1} imply
$\kappa\|LD\|_2<\zeta/2$ and
$|a_\alpha-1|,|a_\beta-1|,|a_\eta-1|<\tau$.
Therefore, by the definition of $\tau$, $\|E\|_2<
\omega\|K\|_2+\frac{\zeta}{2}
+\tau(1+\sqrt{1+\omega^2\|L\|_2^2}+\omega\|L\|_2)
\leq\omega\|K\|_2+\zeta
=\frac{\varpi}{4p}$.

Since $A=A_s+E$, we have
$A^\top P+PA=A_s^\top P+PA_s+E^\top P+PE$.
For any $y$,
$y^\top(E^\top P+PE)y
=2y^\top PEy
\leq2\|P\|_2\|E\|_2\|y\|_2^2$.
Using $\|P\|_2\leq p$, $A_s^\top P+PA_s\preceq-\varpi I$, and bound on
$\|E\|_2$, we obtain
$y^\top(A^\top P+PA)y
\leq-\varpi\|y\|_2^2+2p\|E\|_2\|y\|_2^2
\leq-\frac{\varpi}{2}\|y\|_2^2$.
Thus, $A^\top P+PA\preceq-\frac{\varpi}{2}I$.

Moreover,
$\|A_s\|_2\leq
d_++\sqrt{1+\|C\|_2^2}
\leq d_++\sqrt{5}/2$.
Hence, using bound on
$\|E\|_2$,
$\|A\|_2<d_++\sqrt{5}/2+\varpi/(4p)$.
For convenience, let
$a:=d_++\sqrt{5}/2+\varpi/(4p)$, so that $\|A\|_2<a$. Using $\widehat J=I+\gamma A$,
$A^\top P+PA\preceq-\frac{\varpi}{2}I$, $P\preceq pI$, and
$\|A\|_2<a$, we obtain $\widehat J^\top P\widehat J-P
=\gamma(A^\top P+PA)+\gamma^2A^\top PA\preceq-\gamma(
\frac{\varpi}{2}-\gamma pa^2)I\prec0$, where the last inequality follows from
$\gamma<\varpi/(2pa^2)$. Since $P\succ0$, $\widehat J$ is Schur
stable. Because $\widehat J$ and $J(T)$ are similar, $J(T)$ is also
Schur stable. In particular,
$P_T:=\mathcal T^\top P\mathcal T\succ0$ satisfies
$J(T)^\top P_TJ(T)-P_T\prec0$. 

Having established Schur stability, we next construct the inner
equilibrium and prove local exponential convergence.

First, to establish equilibrium existence, we use the known quantities
$\mathbf b^{\mathrm o}(T)$ and $\delta^{\mathrm o}(T)$, whose
existence is established in Proposition~1. We choose
$\mathbf b^\star(T):=\mathbf b^{\mathrm o}(T)$ as a candidate
equilibrium allocation and then select the corresponding dual
and auxiliary states as follows to verify that all three updates
in (9) leave the candidate state unchanged.

Define
$L_Q:=Q^\top LQ$, which is nonsingular and satisfies
$LQ=QL_Q$ under Assumption~2, and set
$\boldsymbol\lambda^\star(T):=
\frac{\gamma}{\alpha}\delta^{\mathrm o}(T)\mathbf1$ and
$\mathbf z^\star(T):=QL_Q^{-1}Q^\top
(\mathbf c(T)-\mathbf b^{\mathrm o}(T))
+\bar z_0(T)\mathbf1$, where
$\bar z_0(T):=\mathbf1^\top\mathbf z(0;T)/n$.
We now verify that all three DRRA updates leave this candidate
state unchanged. Since $\mathbf1^\top\mathbf c(T)
=\mathbf1^\top\mathbf b^{\mathrm o}(T)=B(T)$,
the difference $\mathbf c(T)-\mathbf b^{\mathrm o}(T)$ belongs
to $\mathbf1^\perp$, so
$QQ^\top(\mathbf c(T)-\mathbf b^{\mathrm o}(T))
=\mathbf c(T)-\mathbf b^{\mathrm o}(T)$.
Using $LQ=QL_Q$ and $L\mathbf1=0$, we obtain
$L\mathbf z^\star(T)=\mathbf c(T)-\mathbf b^{\mathrm o}(T)$.
Hence,
$\mathbf b^\star(T)+L\mathbf z^\star(T)-\mathbf c(T)=\mathbf0$. Moreover, $\mathbf1^\top Q=0$ gives
$\mathbf1^\top\mathbf z^\star(T)
=\mathbf1^\top\mathbf z(0;T)$. The condition (4) gives
$\mathbf g(\mathbf b^{\mathrm o}(T))
=\delta^{\mathrm o}(T)\mathbf1$.
Therefore, using $L\mathbf1=0$,
$(\gamma I-kL)\mathbf g(\mathbf b^\star(T))
-\alpha\boldsymbol\lambda^\star(T)
=\gamma\delta^{\mathrm o}(T)\mathbf1
-\gamma\delta^{\mathrm o}(T)\mathbf1=\mathbf0$.
Also,
$L\boldsymbol\lambda^\star(T)
=(\gamma/\alpha)\delta^{\mathrm o}(T)L\mathbf1=\mathbf0$.
Thus, substitution into (9) verifies that the next
allocation, auxiliary state, and dual state equal their
current values, proving that
$(\mathbf b^\star(T),\mathbf z^\star(T),
\boldsymbol\lambda^\star(T))$ is a DRRA equilibrium.

We next establish local exponential convergence to the equilibrium
and Lyapunov stability of the inner DRRA dynamics.

Let $\widetilde{\mathbf x}^\star(T)$ denote the reduced
equilibrium in the coordinates of Remark~\ref{rem1},
and let $\widetilde\Phi_T$ denote the corresponding update map.

By Assumption 1, $\widetilde\Phi_T$ is continuously differentiable
near $\widetilde{\mathbf x}^\star(T)$, with Jacobian
$D\widetilde\Phi_T(\widetilde{\mathbf x}^\star(T))=J(T)$
given by (10). Using the Schur-stability result established above, define
$H_T:=P_T-J(T)^\top P_TJ(T)\succ0$. Then, for any vector $v$, $\|J(T)v\|_{P_T}^2
=\|v\|_{P_T}^2-v^\top H_Tv
\leq
(1-\frac{\lambda_{\min}(H_T)}{\lambda_{\max}(P_T)})
\|v\|_{P_T}^2$. Since $0\prec H_T\preceq P_T$, choose $q_T\in(0,1)$ such that
$\sqrt{1-\lambda_{\min}(H_T)/\lambda_{\max}(P_T)}
<q_T<1$.
By continuity of $D\widetilde\Phi_T$, there exists $r_T>0$ such
that
$\|D\widetilde\Phi_T(\widetilde{\mathbf x})v\|_{P_T}
\leq q_T\|v\|_{P_T}$ for every $v$ whenever
$\|\widetilde{\mathbf x}
-\widetilde{\mathbf x}^\star(T)\|_{P_T}\leq r_T$.
Since $\mathbf b^\star(T)=\mathbf b^{\mathrm o}(T)$ lies in the
positive interior operating region, $r_T$ can be reduced, if
necessary, so that every state in this neighborhood also has its
allocation component in that region. Define
$\mathcal E_T:=\{\widetilde{\mathbf x}:
\|\widetilde{\mathbf x}
-\widetilde{\mathbf x}^\star(T)\|_{P_T}\leq r_T\}$.
For any $\widetilde{\mathbf x}\in\mathcal E_T$, let
$e:=\widetilde{\mathbf x}
-\widetilde{\mathbf x}^\star(T)$.
Since $\mathcal E_T$ is convex and
$\widetilde\Phi_T(\widetilde{\mathbf x}^\star(T))
=\widetilde{\mathbf x}^\star(T)$, the fundamental theorem of
calculus gives $\widetilde\Phi_T(\widetilde{\mathbf x})
-\widetilde{\mathbf x}^\star(T)
=
\int_0^1
D\widetilde\Phi_T
\bigl(\widetilde{\mathbf x}^\star(T)+se\bigr)e\,ds$. Taking the $P_T$-norm and using the Jacobian bound yields
$\|\widetilde\Phi_T(\widetilde{\mathbf x})
-\widetilde{\mathbf x}^\star(T)\|_{P_T}
\leq q_T\|e\|_{P_T}\leq q_Tr_T<r_T$.
Thus, $\mathcal E_T$ is positively invariant. Let
$V_T(\widetilde{\mathbf x})
:=\|\widetilde{\mathbf x}
-\widetilde{\mathbf x}^\star(T)\|_{P_T}^2$. Since $P_T\succ0$, $V_T$ is positive definite relative to
$\widetilde{\mathbf x}^\star(T)$.
For every $\widetilde{\mathbf x}\in\mathcal E_T$, the preceding
contraction bound gives
$V_T(\widetilde\Phi_T(\widetilde{\mathbf x}))
\leq q_T^2V_T(\widetilde{\mathbf x})$.
Equivalently,
$V_T(\widetilde\Phi_T(\widetilde{\mathbf x}))
-V_T(\widetilde{\mathbf x})
\leq-(1-q_T^2)V_T(\widetilde{\mathbf x})<0$
for every
$\widetilde{\mathbf x}\neq\widetilde{\mathbf x}^\star(T)$
in $\mathcal E_T$. Therefore, if
$\|\widetilde{\mathbf x}(0;T)
-\widetilde{\mathbf x}^\star(T)\|_{P_T}\leq r_T$,
then $\widetilde{\mathbf x}(t;T)\in\mathcal E_T$ for every
$t\geq0$, the allocations remain in the positive interior operating
region, and induction gives
\begin{equation}\label{eq:A-exponential-rate}
\left\|
\widetilde{\mathbf x}(t;T)
-\widetilde{\mathbf x}^\star(T)
\right\|_{P_T}
\leq
q_T^t
\left\|
\widetilde{\mathbf x}(0;T)
-\widetilde{\mathbf x}^\star(T)
\right\|_{P_T}.
\end{equation}
Hence, the reduced equilibrium is locally exponentially stable.
The strict decrease of $V_T$ also establishes Lyapunov stability. Finally, invariance of the auxiliary-state average
(Remark~\ref{rem1}) and
$\mathbf1^\top\mathbf z^\star(T)
=\mathbf1^\top\mathbf z(0;T)$ give
$\mathbf z(t;T)-\mathbf z^\star(T)
=Q(Q^\top\mathbf z(t;T)-Q^\top\mathbf z^\star(T))$.
Since $Q^\top Q=I_{n-1}$, the exponential estimate for the
reduced state also yields exponential convergence of the
full state. 

Consequently, the full equilibrium
$(\mathbf b^\star(T),\mathbf z^\star(T),
\boldsymbol\lambda^\star(T))$
is locally exponentially stable and hence Lyapunov stable
relative to the invariant set
$\{(\mathbf b,\mathbf z,\boldsymbol\lambda):
\mathbf1^\top\mathbf z=\mathbf1^\top\mathbf z^\star(T)\}$.
\end{IEEEproof}

\subsection{Exact Recovery of Centralized Allocation and Utility-Driven Resource-Regeneration Dynamics}\label{sec_3_b}

This subsection establishes that the equilibria of the inner DRRA
dynamics \eqref{eq5}--\eqref{eq7} exactly recover the centralized
optimizer $\mathbf b^{\mathrm o}(T)$ while satisfying the global
resource constraint \eqref{eq1}. Furthermore, it demonstrates that,
after inner convergence, the local resource-reference updates
\eqref{eq8} reproduce the centralized regeneration dynamics \eqref{eq2}.

Lemma~\ref{lem1} establishes resource feasibility by proving
that the equilibrium allocation satisfies the resource constraint \eqref{eq1}.


\begin{lemma}
\label{lem1}
Given Assumptions~\ref{ass1} and~\ref{ass2}, fix a regeneration
round $T$. When the inner allocation converges to $\mathbf b^\star(T)$
under the conditions of Proposition~\ref{pro2},
it satisfies the global resource constraint \eqref{eq1}, i.e.,
$\sum_{i=1}^{n}b_i^\star(T)=\sum_{i=1}^{n}c_i(T)=B(T)$.
\end{lemma}

\begin{IEEEproof}
At the equilibrium established in Proposition~\ref{pro2},
the dynamics \eqref{eq7} satisfy
$\lambda_i(t+1;T)=\lambda_i(t;T)$. Since $\beta>0$, the corresponding
update term in \eqref{eq7} must vanish, yielding
\begin{equation}
\label{eq11}
b_i^\star(T)
+
\sum_{j\in\mathcal N_i}
a_{ij}
\left(
z_i^\star(T)-z_j^\star(T)
\right)
-
c_i(T)
=
0,
\;
\forall i\in\mathcal V.
\end{equation}
\indent By the definition of the weighted graph Laplacian, the coupling term in
\eqref{eq11} satisfies
$\sum_{j\in\mathcal N_i}a_{ij}(z_i^\star(T)-z_j^\star(T))
=(L\mathbf z^\star(T))_i$. Summing
\eqref{eq11} over all agents gives
$\sum_{i=1}^{n}b_i^\star(T)+\mathbf 1^\top L\mathbf z^\star(T)
-\sum_{i=1}^{n}c_i(T)=0$. Since the graph is weight-balanced by Assumption~\ref{ass2}, its
Laplacian satisfies $\mathbf 1^\top L=\mathbf 0^\top$. Hence,
$\mathbf 1^\top L\mathbf z^\star(T)=0$, implying
$\sum_{i=1}^{n}b_i^\star(T)=\sum_{i=1}^{n}c_i(T)$. Finally, using $\sum_{i=1}^{n}c_i(T)=B(T)$ yields
$\sum_{i=1}^{n} b_i^\star(T) = \sum_{i=1}^{n} c_i(T) = B(T)$.
\end{IEEEproof}


Lemma~\ref{lem2} guarantees consensus of the local dual
variables governed by \eqref{eq7} at equilibrium.


\begin{lemma}\label{lem2}
Given Assumptions~\ref{ass1} and~\ref{ass2}, fix a
regeneration round $T$. When the inner dynamics \eqref{eq5}-\eqref{eq7} converge to an
equilibrium $(\mathbf b^\star(T),\mathbf z^\star(T),\boldsymbol{\lambda}^\star(T))$ under the conditions of Proposition~\ref{pro2},
the equilibrium dual variables satisfy
\begin{equation}
\label{eq12}
\lambda_1^\star(T)
=
\lambda_2^\star(T)
=
\cdots
=
\lambda_n^\star(T)
=:
\lambda_c^\star(T),
\end{equation}
where $\lambda_c^\star(T)\in\mathbb{R}$ denotes the common dual value.
\end{lemma}

\begin{IEEEproof}
At the equilibrium established in Proposition~\ref{pro2},
the auxiliary dynamics \eqref{eq6} satisfy
$z_i(t+1;T)=z_i(t;T)$. Since $\eta>0$, the update term in
\eqref{eq6} must vanish, yielding
\begin{equation}
\label{eq13}
\sum_{j\in\mathcal N_i}
a_{ij}
\left(
\lambda_i^\star(T)-\lambda_j^\star(T)
\right)
=
0,
\qquad
\forall i\in\mathcal V.
\end{equation}
\indent By the definition of the weighted graph Laplacian, the coupling term in
\eqref{eq13} satisfies
$\sum_{j\in\mathcal N_i}a_{ij}
\left(
\lambda_i^\star(T)-\lambda_j^\star(T)
\right)
=
(L\boldsymbol{\lambda}^\star(T))_i$.
Collecting the equations for all agents gives
$L\boldsymbol{\lambda}^\star(T)=\mathbf 0$. Since $\mathcal G$ is strongly connected by Assumption~\ref{ass2}, the right nullspace of its Laplacian $L$ is one-dimensional. Moreover, $L\mathbf 1=\mathbf 0$. Hence,
$\ker(L)=\operatorname{span}\{\mathbf 1\}$. Therefore,
$\boldsymbol{\lambda}^\star(T)\in\ker(L)$ implies that there exists a scalar
$\lambda_c^\star(T)\in\mathbb R$ such that
$\boldsymbol{\lambda}^\star(T)=\lambda_c^\star(T)\mathbf 1$, which proves
\eqref{eq12}.
\end{IEEEproof}


Using Lemma~\ref{lem1} and~\ref{lem2}, the following theorem establishes exact recovery of the centralized optimal resource allocation at an equilibrium in the positive interior operating region.


\begin{theorem}
\label{thm1}
Under Assumptions~\ref{ass1} and~\ref{ass2},
using Lemmas~\ref{lem1} and~\ref{lem2}
and the conditions of Proposition~\ref{pro2},
fix a regeneration round $T$. Let $\lambda_c^\star(T)$ be the common dual value in \eqref{eq12} and define
$\bar\delta(T):=\frac{\alpha}{\gamma}\lambda_c^\star(T)$. Then the marginal utilities at the equilibrium allocation $\mathbf b^\star(T)$ (local convergence and stability guaranteed by Proposition~\ref{pro2}) satisfy
\begin{equation}
\label{eq14}
\frac{dU_i\!\left(b_i^\star(T)\right)}{db_i}
=
\bar{\delta}(T),
\;
\forall i\in\mathcal V.
\end{equation}
Moreover, $\mathbf b^\star(T)=\mathbf b^{\mathrm{o}}(T)$.
Consequently, the DRRA equilibrium satisfies the KKT conditions of the centralized CRRA problem in Proposition~\ref{pro1}, thereby recovering the unique centralized optimal allocation.
\end{theorem}

\begin{IEEEproof}
Define the equilibrium marginal utility of agent $i$ as
$g_i^\star(T)\!:=\!
\frac{dU_i\!\left(b_i^\star(T)\right)}{db_i}$,
and let
$\mathbf g^\star(T):=
[g_1^\star(T),\ldots,g_n^\star(T)]^\top$. At the steady state of the primal resource-allocation dynamics \eqref{eq5}, we have
$b_i(t+1;T)=b_i(t;T)$. Hence $\forall i\in\mathcal V$,
\begin{equation}
\label{eq15}
\gamma g_i^\star(T)
\!+\!
k\sum_{j\in\mathcal N_i}\!
a_{ij}
\left(
g_j^\star(T)\!-\!g_i^\star(T)
\right)
\!-\!
\alpha\lambda_i^\star(T)
\!=
0.
\end{equation}
\indent By Lemma~\ref{lem2},
$\lambda_i^\star(T)=\lambda_c^\star(T)$ for all
$i\in\mathcal V$. Moreover, by the definition of the weighted graph Laplacian,
$
\sum_{j\in\mathcal N_i}
a_{ij}
\left(
g_j^\star(T)-g_i^\star(T)
\right)
=
-(L\mathbf g^\star(T))_i.
$
Stacking \eqref{eq15} over all agents yields
$
(\gamma I-kL)\mathbf g^\star(T)
=
\alpha\lambda_c^\star(T)\mathbf 1
$. Since $L\mathbf 1=\mathbf 0$ by Assumption~\ref{ass2},
$(\gamma I-kL)\mathbf 1=\gamma\mathbf 1$. 

Moreover, the Schur stability of $J(T)$ in Proposition~\ref{pro2} excludes $\gamma/k\in\sigma(L)$. Suppose instead that $(\gamma I-kL)\mathbf v=\mathbf 0$ for some $\mathbf v\neq\mathbf 0$. Since $(\gamma I-kL)\mathbf 1=\gamma\mathbf 1$, $\mathbf v\notin\operatorname{span}\{\mathbf 1\}$. Let $\mathbf p=\mathbf v+\theta\mathbf 1$ with $\theta=-\mathbf 1^\top D(T)^{-1}\mathbf v/(\mathbf 1^\top D(T)^{-1}\mathbf 1)$ and $\mathbf u=D(T)^{-1}\mathbf p\neq\mathbf 0$. Since $\mathbf 1^\top\mathbf u=0$ and $\operatorname{range}(LQ)=\operatorname{range}(L)=\{\mathbf w:\mathbf 1^\top\mathbf w=0\}$ by Assumption~\ref{ass2}, there exists $\mathbf y$ with $LQ\mathbf y=-\mathbf u$. Then $(\mathbf u,\mathbf y,-(\theta\gamma/\alpha)\mathbf 1)$ is an eigenvector of $J(T)$ in \eqref{eq10} with eigenvalue $1$, which contradicts Schur stability.
Hence, $\gamma/k$ is not an eigenvalue
of $L$, implying that $\gamma I-kL$ is nonsingular. Therefore,
$(\gamma I-kL)^{-1}\mathbf 1=\frac{1}{\gamma}\mathbf 1$. Applying this identity to $
(\gamma I-kL)\mathbf g^\star(T)
=
\alpha\lambda_c^\star(T)\mathbf 1
$ gives $\mathbf g^\star(T)
=
\frac{\alpha}{\gamma}\lambda_c^\star(T)\mathbf 1$.

Using the definition of $\bar{\delta}(T)$ from the theorem statement,
\eqref{eq14} follows directly, which is the KKT stationarity condition \eqref{eq4} with multiplier $\bar{\delta}(T)$. By Lemma~\ref{lem1},
$\sum_{i=1}^{n}b_i^\star(T)=B(T)$. Therefore,
$\mathbf b^\star(T)$ satisfies both the KKT stationarity condition \eqref{eq4} and the resource constraint \eqref{eq1}. Since the objective function is strictly concave and the feasible set of \eqref{eq1} is convex, these KKT conditions uniquely characterize the centralized optimum. Therefore,
$\mathbf b^\star(T)=\mathbf b^{\mathrm{o}}(T)$.
\end{IEEEproof}

The following corollary establishes that the local
resource-reference updates \eqref{eq8} exactly recover the
centralized resource-regeneration dynamics \eqref{eq2}.

\begin{corollary}
\label{cor4}
Under the conditions of Proposition~\ref{pro2} and Theorem~\ref{thm1}, fix a regeneration round $T$. The local resource-reference dynamics \eqref{eq8} of DRRA exactly reproduce the centralized resource-regeneration dynamics \eqref{eq2}.
\end{corollary}

\begin{IEEEproof}
By Proposition~\ref{pro2},
the inner dynamics \eqref{eq5}--\eqref{eq7} converge locally
exponentially to a Lyapunov-stable equilibrium. After convergence,
summing \eqref{eq8} over all agents gives
$\sum_{i=1}^{n}c_i(T+1)
\!=\!\sum_{i=1}^{n}c_i(T)
\!+\!\epsilon\sum_{i=1}^{n}U_i(b_i^\star(T))$.
By Theorem~\ref{thm1},
$\mathbf b^\star(T)\!=\!\mathbf b^{\mathrm o}(T)$.
Therefore, the total resource increment generated by the
local updates is
$\sum_{i=1}^{n}(c_i(T+1)-c_i(T))
\!=\!\epsilon\sum_{i=1}^{n}U_i(b_i^{\mathrm o}(T))$,
which is exactly the utility-driven increment in \eqref{eq2}.
Since $\sum_{i=1}^{n}c_i(T)\!=\!B(T)$, this yields
$\sum_{i=1}^{n}c_i(T+1)
\!=\!B(T)\!+\!\epsilon\sum_{i=1}^{n}U_i(b_i^{\mathrm o}(T))
\!=\!B(T+1)$.
Thus, the local updates \eqref{eq8} exactly reproduce the
centralized resource-regeneration dynamics \eqref{eq2}.
\end{IEEEproof}


\begin{figure}[t]
\centering
\includegraphics[width=\columnwidth]{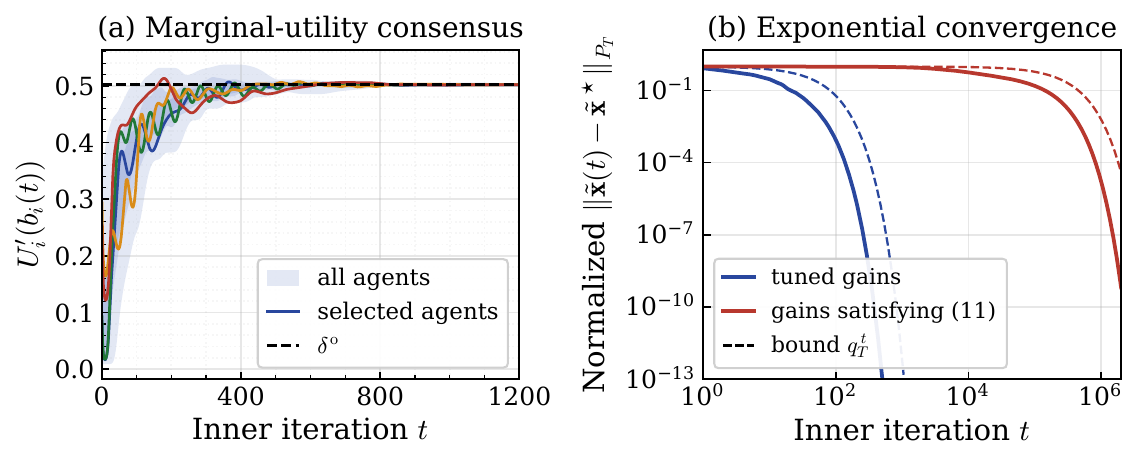}
\vspace{-6mm}
\caption{(a)~Marginal utilities in the first round of a representative ER instance ($n=20$) for four selected agents (lines) and all agents (shading), with $\delta^{\mathrm o}$ (dashed). (b)~Normalized $P_T$-norm inner error after the first outer update for tuned gains and gains satisfying \eqref{eqa1}, with the bounds $q_T^t$ (dashed).}
\label{fig:main}
\vspace{-2mm}
\end{figure}

\begin{figure}[t]
\centering
\includegraphics[width=\columnwidth]{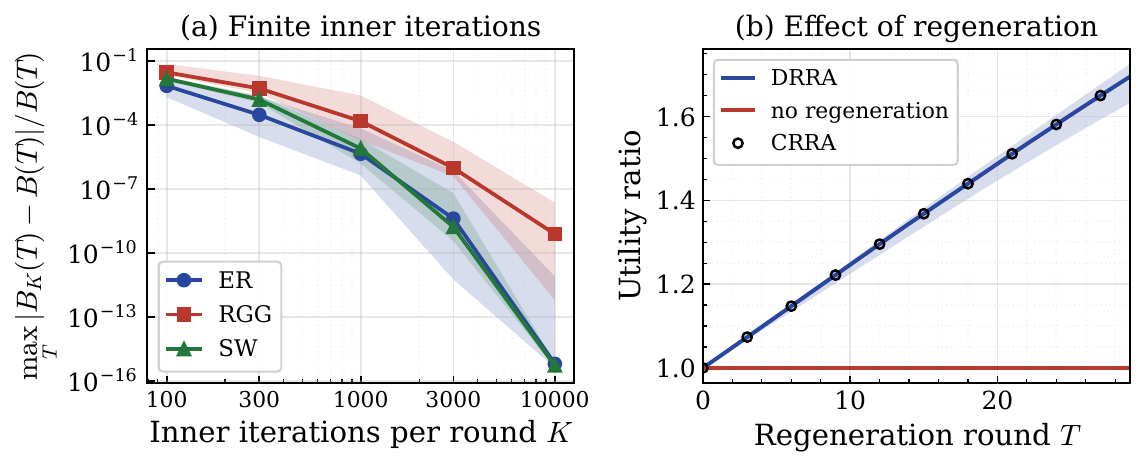}
\vspace{-6mm}
\caption{(a)~Relative deviation of the total resource $B_K(T)$, obtained with $K$ inner iterations per round, from \eqref{eq2} (median and interquartile range per topology). (b)~Aggregate utility per round relative to the case without regeneration (median and 10--90\% range), with the CRRA benchmark (circles).}
\label{fig:cert}
\vspace{-4mm}
\end{figure}

\section{Numerical Study}\label{sec_5}

We evaluate DRRA through paired Monte Carlo simulations, in which DRRA, distributed resource allocation without regeneration, and the CRRA benchmark are applied to the same randomly generated instances. Each instance has $n\in\{20,30\}$ agents connected by an undirected Erd\H{o}s--R\'enyi (ER) graph with edge probability $0.2$, a random geometric graph (RGG) with radius $0.38$ in the unit square, or a Watts--Strogatz small-world (SW) graph with four nearest neighbors and rewiring probability $0.2$. Disconnected realizations are discarded, and all edges have unit weight, so every graph satisfies Assumption~\ref{ass2}. Six instances are generated for each topology and size, giving $36$ instances. Each agent is randomly assigned either $U_i(b)\!=\!w_i(1-e^{-a_ib})$ or $U_i(b)\!=\!w_i\ln(1+a_ib)$, with $w_i$ and $a_i$ drawn uniformly from $[0.8,1.6]$ and $[0.5,1.2]$, respectively, and the utilities are resampled until the CRRA optimizer is interior (Assumption~\ref{ass1}). We set $B(0)\!=\!n$, $\epsilon\!=\!0.05$, $c_i(0)\!=\!b_i(0)\!=\!1$, and $z_i(0)\!=\!\lambda_i(0)\!=\!0$, and simulate $30$ rounds. In each round, the inner dynamics \eqref{eq5}--\eqref{eq7} run with $\mathbf c(T)$ fixed until successive iterates differ by less than $10^{-12}$ in every component. Each agent then applies \eqref{eq8}, and the next round starts from the final inner state. The CRRA optimum is computed by bisection on the dual variable. For each network, the gains are tuned offline, using centralized knowledge of $L$ and the utilities, to minimize the larger of the two spectral radii of $J(T)$ in \eqref{eq10} at the CRRA optima for $B(0)$ and $B(30)$ through a random search followed by Nelder--Mead refinement. The spectral radius of $J(T)$ remained below $0.9999$ in every round of every instance, so the condition of Proposition~\ref{pro2}, and hence $\gamma/k\notin\sigma(L)$, held throughout. Every converged allocation was positive. The transient inner iterates also remained positive, except in the first round of one instance, which starts outside the local region of Proposition~\ref{pro2} and in which one iterate briefly reached $-0.044$.

Fig.~\ref{fig:main}(a) shows the first round of a representative ER instance with $n\!=\!20$, in which all marginal utilities converge to the common multiplier $\delta^{\mathrm o}$, as required by \eqref{eq4}. Fig.~\ref{fig:main}(b) illustrates Proposition~\ref{pro2} on a denser ER graph ($n\!=\!20$, edge probability $0.6$) after the first outer update, starting from the previous equilibrium. For this graph and the curvature matrix $D(T)$ at the new equilibrium, the bounds \eqref{eqa1} require $0\!<\!\gamma\!<\!1.08\times10^{-3}$, $0\!<\!k\!<\!9.41\times10^{-5}\gamma$, $|\alpha-\gamma|,|\beta-\gamma|\!<\!\tau\gamma$, and $|\eta-\omega^2\gamma|\!<\!\tau\omega^2\gamma$, where $\tau\!=\!3.13\times10^{-4}$ and $\omega^2\!=\!8.62\times10^{-4}$ are computed from the
definitions preceding \eqref{eqa1} in the proof of
Proposition~\ref{pro2}. We compare the gains $\gamma\!=\!9.75\times10^{-4}$, $k\!=\!4.59\times10^{-8}$, $\alpha\!=\!\beta\!=\!\gamma$, and $\eta\!=\!\omega^2\gamma$, which satisfy \eqref{eqa1}, with the tuned gains $(\gamma,k,\alpha,\eta,\beta)\!=\!(3.63,1.75\times10^{-2},3.39,4.09\times10^{-3},9.90\times10^{-2})$, which satisfy only the spectral condition of Proposition~\ref{pro2}. In both cases, $P_T$ is computed from $(J(T)/q_T)^\top P_T(J(T)/q_T)-P_T\!=\!-I$ with $q_T\!=\!(1+\rho_T)/2$, where $\rho_T$ is the spectral radius of $J(T)$. This $P_T$ satisfies the LMI of Proposition~\ref{pro2}, and the measured error of the nonlinear inner dynamics in the $P_T$-norm, normalized by its initial value, remains below $q_T^t$, with $q_T\!=\!0.999995$ and $q_T\!=\!0.972$, respectively, until it reaches double-precision accuracy. The gains satisfying \eqref{eqa1} reduce the error only to $6.6\times10^{-10}$ after $2\times10^{6}$ iterations, whereas the tuned gains reach $10^{-10}$ within $380$ iterations, showing that \eqref{eqa1} is sufficient but conservative. In this example,
centralized offline tuning reduces online communication rounds,
while the DRRA updates remain distributed.

Fig.~\ref{fig:cert}(a) quantifies the effect of a finite number $K$ of inner iterations per round. Here, $B_K(T)$ is obtained by stopping the inner dynamics after $K$ iterations in each round and applying \eqref{eq8} to the resulting allocations, and its deviation from the centralized dynamics \eqref{eq2} is measured by $\max_T|B_K(T)-B(T)|/B(T)$ over the $30$ rounds. This deviation decreases approximately geometrically with $K$, with a median over all instances of $1.6\times10^{-2}$ at $K\!=\!100$ and $7.9\times10^{-16}$ at $K\!=\!10^4$. With the tolerance-based stopping rule, the median number of inner iterations per round is $3954$ (ER), $11065$ (RGG), and $4087$ (SW). The largest deviations over all instances and rounds are then $|\mathbf 1^\top\mathbf c(T)-B(T)|\!=\!6.2\times10^{-11}$ and $\|\mathbf b^\star(T)-\mathbf b^{\mathrm o}(T)\|_\infty\!=\!1.4\times10^{-9}$, consistent with Theorem~\ref{thm1} and Corollary~\ref{cor4}.

Fig.~\ref{fig:cert}(b) compares DRRA with distributed resource allocation without regeneration, obtained by setting $\epsilon\!=\!0$ in \eqref{eq8} for the same instances and inner dynamics. In the final round, regeneration raises the aggregate utility to $1.70$ times its value without regeneration (median), and the cumulative utility over $30$ rounds increases by $34.9\%$ on average. The DRRA utilities match the CRRA benchmark in every round, with relative deviation below $10^{-10}$.

\section{Conclusion} \label{sec_6}
This letter established a DRRA framework for utility-driven resource regeneration in multi-agent networks. We proved exact recovery of the centralized optimal allocation and preservation of the global resource constraint. Under sufficient gain conditions, the inner dynamics converge
locally exponentially to a stable equilibrium, and subsequent updates exactly recover centralized resource regeneration,
as supported by simulations.

\bibliographystyle{IEEEtran}
\bibliography{references}

\end{document}